\documentclass[a4paper,11pt,leqno]{article}
\usepackage{amsmath,amssymb}
\usepackage{amsfonts}
\usepackage{eucal}
\usepackage{amsthm}

\newcommand{\bigpare}[1]{\bigl(#1\bigr)}
\newcommand{\biggpare}[1]{\biggl(#1\biggr)}

\newcommand{\bigbrac}[1]{\bigl[#1\bigr]}

\newcommand{\biggbrac}[1]{\biggl[#1\biggr]}

\newcommand{\jap}[1]{\langle #1 \rangle}

\def\b{\beta}
\def\c{\gamma}
\def\d{\delta}
\def\e{\varepsilon}
\def\f{\varphi}

\def\l{\lambda}

\def\s{\sigma}

\def\x{\xi}
\def\y{\eta}

\newcommand{\D}{\Delta}

\def\re{\mathbb{R}}

\def\pa{\partial}

\newcommand{\supp}{\mathrm{{supp}}}

\newtheorem{thm}{Theorem}

\newtheorem{cor}[thm]{Corollary}

\theoremstyle{definition}

\newtheorem{ass}{Assumption}

\newtheorem{example}{Example}

\theoremstyle{remark}
\newtheorem{rem}{Remark}

\title{A Remark on the Mourre Theory for Two Body Schr\"odinger operators\footnote{2000 
Mathematical Subject Classification: 81U05, 35P25, 47A40. 
Keywords: scattering theory, Schr\"odinger operators, Mourre estimate.}}
\author{Shu N{\sc akamura}%
\footnote{Graduate School of Mathematical Sciences, 
University of Tokyo, 3-8-1 Komaba, Meguro Tokyo, 
153-8914 Japan. 
E-mail: {\tt shu@ms.u-tokyo.ac.jp}.  
Partially supported by JSPS Grant Kiban (A) 21244008.} }

\begin{document}
\maketitle

\begin{abstract}
On this short note, we apply the Mourre theory of the limiting absorption with 
{\it difference}\/ type conditions on the potential, instead of conditions on the derivatives. 
In order that we modify the definition of the conjugate operator, and we apply the standard 
abstract Mourre theory. We also discuss examples to which the method applies. 
\end{abstract}


\section{Introduction}

We consider the Schr\"odinger operator on $\re^d$, i.e, 
\[
H= H_0+V(x), \quad H_0=-\frac12 \sum_{j=1}^d \frac{\pa^2}{\pa x_j^2}\quad
\text{on }\mathcal{H}=L^2(\re^d)
\]
with $d\geq 1$. $V(x)$ is the potential, and we always suppose $V(x)$ is a real-valued locally 
$L^2$-function. 

Let $I\subset \re$ be an open interval. We say {\it the Mourre theory applies to $H$ on $I$}, 
if for any interval $J\Subset I$, there is a self-adjoint operator $A$ on $\mathcal{H}$ such that 
\begin{enumerate}
\renewcommand{\labelenumi}{(\roman{enumi})} 
\item For $z\in\rho(H)$, $t\mapsto e^{itA} (H-z)^{-1} e^{-itA}$ is a $\mathcal{L}(\mathcal{H})$-valued 
$C^2$-function on $\re$; 
\item There is $c>0$ such that 
\begin{equation}
E_J(H) [H,iA] E_J(H)\geq c E_J(H)+K
\end{equation}
with some compact operator $K$, where $E_J(H)$ is the spectral projection. 
\end{enumerate}
It is well-known (see e.g., \cite{ABG, DG}) that under these conditions, the following properties hold:

\begin{enumerate}
\renewcommand{\labelenumi}{(\alph{enumi})}
\item $\s_p(H)\cap I$ is discrete, and each eigenvalues are of finite rank;
\item $H$ is absolutely continuous on $I\setminus\s_p(H)$;
\item Let $\c>1/2$. Then for each $\l\in I\setminus \s_p(H)$, 
\[
\lim_{\e\downarrow 0}\jap{A}^{-\c}(H-\l\pm i\e)^{-1}\jap{A}^{-\c}\in \mathcal{L}(\mathcal{H})
\]
exist and the limits are H\"older continuous in $\l\in I\setminus\s_p(H)$. 
\end{enumerate}

Here we have used the standard notation: $\jap{x}=(1+|x|^2)^{1/2}$. 

When we apply the Mourre theory to 2 body Schr\"odinger operators, we usually use 
$A=\frac{1}{2i}(x\cdot\pa_x+\pa_x\cdot x)$, and some derivative conditions are imposed on the potential, 
at least on the long-range part. 
Instead, we suppose difference type conditions as follows. 
Let $\b>0$, and let $e_j=(\d_{jk})_{k=1}^d\in\re^d$ 
($j=1,\dots, d$) be the standard basis of $\re^d$. We set 
\[
T_j^\b f(x)= f(x+\b e_j), \quad T_j^{\b*}f(x)= f(x-\b e_j), 
\]
 and also 
\[
\D_j^\b f(x)= \frac{1}{\b}(T_j^\b-1)f(x)= \frac{1}{\b}(f(x+\b e_j)-f(x)), 
\]
for a function $f$ on $\re^d$, $x\in\re^d$ and $j=1,\dots, d$. 

\begin{ass}
Let $\b>0$. $V(x)$ and $x_j \D_j^\b V(x)$ ($j=1,\dots, d$) are $H_0$-compact. Moreover, 
$x_j x_k \D_j^\b \D_k^\b V(x)$ ($j,k=1,\dots, d$) are $H_0$-bounded. 
\end{ass}

\begin{thm}
Suppose Assumption~A with $\b>0$, and let $I=\bigpare{0,\tfrac12(\pi/\b)^2}$. 
Then the Mourre theory applies to $H$ on $I$. Hence, in particular, the properties (a)--(c) holds on $I$. 
\end{thm}

\begin{cor}
Suppose Assumption~A holds for all $\b>0$. Then the Mourre theory applies to $H$ on $(0,\infty)$. 
\end{cor}

\begin{rem}
In Corollary~2, we do not assume Assumption~A with uniform bounds in $\b>0$. 
Hence, $V$ is not necessarily differentiable. 
\end{rem}

\begin{example}
Suppose $V=V_1+V_2+V_3$, where 
\begin{enumerate}
\renewcommand{\labelenumi}{(\roman{enumi})}
\item $|x|^2 V(x)$ is $H_0$-bounded. 
\item $V_2\in C^1(\re^d)$. $V_2$ and $|x|^2\pa_{x_j} V_2$ ($j=1,\dots, d$) are $H_0$-compact. 
\item $V_3\in C^2(\re^d)$. $V_3$ and $x_j\pa_{x_j} V_3$ ($j=1,\dots, d$) are $H_0$-compact, and 
$x_j x_k \pa_{x_j}\pa_{x_k} V_3$ ($j,k=1,\dots, d$) are $H_0$-bounded. 
\end{enumerate}
Then $V$ satisfies Assumption~A with any $\b>0$. This is a variation of the standard assumption of the 
Mourre theory for 2 body Schr\"odinger operators. 
\end{example}

\begin{example}
Suppose $W(x)$ is a $\b$-periodic locally $L^p$-function, i.e., 
\[
W(x+\b e_j)= W(x), \quad x\in\re^d, \ j=1,\dots, d,
\]
where $p=2$ if $d\leq 3$ and $p>d/2$ if $d\geq 4$. 
Let $\c>0$ and we set 
\[
V(x)=\jap{x}^{-\c} W(x).
\]
Then $V(x)$ satisfies Assumption~A with the above $\b$, and hence $H$ is 
absolutely continuous except for discrete eigenvalues on $\bigpare{0,\tfrac12(\pi/\b)^2}$. 
This example shows that even the long-range part may be rather singular for the Mourre theory
to be applied. 
\end{example}

The Mourre theory is one of the most useful method in the scattering theory \cite{Mo}. 
For comprehensive reviews and applications, see, for example, \cite{ABG}, \cite{BS}, \cite{DG}, 
\cite{Ge}, \cite{T}, \cite{Y}, etc. We use the formulation due to G\'erard \cite{Ge}. 
Usually a differential operator (the dilation generator, in particular) is used as the conjugate operator 
$A$, and hence some discussion about the differentiability of the potential is necessary, 
though it is possible to avoid differentiability assumptions using approximation arguments. 
We employ a conjugate operator which is closely related to the difference operator, and 
this is partially motivated by the Mourre theory for difference operator (\cite{BS}, \cite{IK}).


\section{Proof}

We fix $\b>0$ and suppose Assumption~A in the following. 
We denote the Fourier transform by $\mathcal{F}$: 
\[
\mathcal{F}\f(\x)= (2\pi)^{-d/2} \int e^{-ix\cdot\x}\f(x)dx, \quad\x\in\re^d, \f\in\mathcal{S}(\re^d).
\]
We write 
\[
Q_ju(x)=\frac{1}{2i\b} (T_j^\b -T_j^{\b*})u(x)= \frac{1}{2i\b} (u(x+\b e_j)-u(x-\b e_j)).
\]
We note 
\[
\mathcal{F} Q_j \mathcal{F^*}u(\x)= \frac{1}{\b}\sin (\b\x_j)u(\x), \quad \x\in\re^d, u\in L^2(\re^d). 
\]
We now define 
\[
Au= \frac12 \sum_{j=1}^d \bigpare{Q_j x_j +x_j Q_j}u\quad \text{for }u\in \mathcal{S}(\re^d). 
\]
We then note 
\[
-i\mathcal{F}A\mathcal{F}^* =\frac{1}{2\b} \sum_{j=1}^d \biggpare{\sin(\b\x_j)\frac{\pa}{\pa\x_j}+
\frac{\pa}{\pa\x_j}\sin(\b\x_j)}
\]
is a first order differential operator which generates a unitary group through a change of coordinates. 
This implies, in particular, $\mathcal{F} A \mathcal{F}^*$ is essentially self-adjoint on $\mathcal{S}(\re^d)$, 
and hence $A$ is also essentially self-adjoint on $\mathcal{S}(\re^d)$. Moreover, $e^{-itA}$ leaves 
the domain of $H_0$ and $H$ invariant. 

Now by easy computations, we have 
\[
\mathcal{F}[H_0, iA]\mathcal{F}^* = \sum_{j=1}^d \frac{1}{\b} \sin(\b \x_j)\x_j,
\]
and it is easy to see 
\[
\sum_{j=1}^d \frac{1}{\b} \sin(\b \x_j)\x_j>0  \quad \text{if}\quad  0<|\x|< \frac{\pi}{\b}.
\]
We let $\y>0$ sufficiently small and choose $f\in C_0^\infty(\re)$ such that 
\[
\supp\, f\subset \biggbrac{\frac{\y}{2}, \frac12 \biggpare{\frac{\pi}{\b}-\frac{\y}{2}}^2}; 
\quad f(t)=1\text{ if } t\in \biggbrac{\y,\frac12\biggpare{\frac{\pi}{\b}-\y}^2}.
\]
Then we learn 
\[
f(H_0)[H_0,iA]f(H_0) \geq \d f(H_0)^2
\]
with some $\d>0$. 

It is well-known that $f(H)-f(H_0)$ is compact if $V$ is $H_0$-compact, 
and by using the standard argument, we have 
\begin{equation}
f(H) [H_0,iA]f(H) \geq \d f(H)^2 +K_1
\end{equation}
with a compact operator $K_1$. 

Next we consider $[V, iA]$. By straightforward computation, we have 
\begin{equation}
A= \frac{1}{4i\b}\sum_{j=1}^d (x_j T_j^\b-x_j T_j^{\b*}) 
+\frac{1}{4i}\sum_{j=1}^d (T_j^\b +T_j^{\b*}).
\end{equation}
The second sum in the right hand side is bounded and commutes with $H_0$, 
and hence its commutator with $V$ is $H_0$-compact by the assumption. We also have 
\begin{align*}
[T_j^\b,V] u(x) &= T_j^\b V u(x)- V T_j^\b u(x) \\
&= V(x+\b e_j) u(x+\b e_j) - V(x) u(x+\b e_j) \\
&= (V(x+\b e_j)-V(x)) u(x+\b e_j) \\
&=(\b\Delta_j^\b V) T_j^\b u(x). 
\end{align*}
This implies 
\[
[x_j T_j^\b, V] = x_j [T_j^\b, V] = \b(x_j \Delta_j^\b V)T_j^\b 
\]
is $H_0$-compact again by the assumption. Similarly, we can show $[x_j T_j^*, V]$ is 
$H_0$-compact:
\[
[x_j T_j^{\b*},V]= x_j T_j^{\b*}(\b\Delta_j^\b V) 
= \b T_j^{\b*} (x_j\Delta_j^\b V)+\b T_j^{\b*} (\Delta_j^\b V). 
\] 
Thus we learn $[V, iA]$ is $H_0$-compact. Combining this 
with (2), we obtain the Mourre inequality (1). 

It remains to show $e^{itA} (H-z)^{-1} e^{-itA}$ is a $C^2$-class function in $t$. 
Since $e^{-itA}$ leaves $H^2(\re^d)= D(H)=D(H_0)$ invariant, it suffices to show 
$[H, iA]$ and $[[H, iA], iA]$ are $H_0$-bounded. By the above expressions of $[H_0,iA]$ 
and $[V,iA]$, $[H, iA]$ is obviously $H_0$-bounded. $[[H_0,iA],iA]$ is computed as 
\[
\mathcal{F} [[H_0,iA],iA]\mathcal{F}^* = \frac{1}{\b^2}\sum_{j=1}^d \biggpare{
\sin(\b\x_j)\frac{\pa}{\pa\x_j}\bigbrac{\sin(\b\x_j)\x_j}}\cdot 
\]
and hence it is $H_0$-bounded. $[[V,iA],iA]$ can be computed and estimated using (3) as above. 
For example, we have 
\begin{align*}
[[V, x_j T_j^\b], x_k T_k^\b] &= \b [x_k T_k^\b,(x_j\Delta_j^\b V)T_j^\b] \\
&= \b x_k [T_k^\b,  (x_j\Delta_j^\b V) ] T_j^\b
+\b x_j (\Delta_j^\b V)[x_k, T_j^\b] T_k^\b\\
&= \b^2 x_jx_k (\Delta_j^\b\Delta_k^\b V) T_j^\b T_k^\b \\
&\quad +\b \d_{jk}x_k T_k^\b (\Delta_j^\b V) T_j^\b -\b \d_{jk} x_j (\Delta_j^\b V) T_j^\b T_k^\b\\
&= \b^2 [x_jx_k (\Delta_j^\b\Delta_k^\b V)] T_j^\b T_k^\b \\
&\quad +\b \d_{jk} T_k^\b [x_j (\Delta_j^\b V)] T_j^\b -\b \d_{jk} T_k^\b(\Delta_j^\b V) T_j^\b\\
&\quad -\b \d_{jk} [x_j (\Delta_j^\b V)] T_j^\b T_k^\b
\end{align*}
and each term is $H_0$-bounded by the assumption. Other terms in the expansion of $[[V, iA],iA]$ 
can be computed similarly. \qed 


\end{document}